# Fractal Symbolic Analysis for Program Transformations

Nikolay Mateev, Vijay Menon, Keshav Pingali
Department of Computer Science,
Cornell University, Ithaca, NY 14853

## Abstract

Restructuring compilers use dependence analysis to prove that the meaning of a program is not changed by a transformation. A well-known limitation of dependence analysis is that it examines only the memory locations read and written by a statement, and does not assume any particular interpretation for the operations in that statement. Exploiting the semantics of these operations enables a wider set of transformations to be used, and is critical for optimizing important codes such as LU factorization with pivoting.

Symbolic execution of programs enables the exploitation of such semantic properties, but it is intractable for all but the simplest programs. In this paper, we propose a new form of symbolic analysis for use in restructuring compilers. *Fractal symbolic analysis* compares a program and its transformed version by repeatedly simplifying these programs until symbolic analysis becomes tractable, ensuring that equality of simplified programs is sufficient to guarantee equality of the original programs. We present a prototype implementation of fractal symbolic analysis, and show how it can be used to optimize the cache performance of LU factorization with pivoting.

## 1 Introduction

Modern compilers perform source-level transformations of programs to enhance locality and parallelism. Before such transformations can be performed, the source program must be analyzed to ensure that the proposed transformation does not violate the semantics of the program. The most commonly used analysis technique is *dependence analysis*. The goal of this technique is to compute a partial order between the statements of the program such that any statement reordering consistent with this partial order is guaranteed to leave the output

[0]This work was supported by NSF grants CCR-9720211, EIA-9726388 and ACI-9870687.

```
S1: a = 2*a;                    a = a+b;
S2: b = 2*b;         =>         b = 2*b;
S3: a = a+b;                    a = 2*a;
(a) Original program       (b) Transformed program
```

Figure 1: Simple Reordering of Statements

of the program unchanged. In general, three kinds of dependences may exist from a statement S1 to a statement S2 executed after it.

1. *Flow-dependence:* A flow-dependence exists from S1 to S2 if S1 may write to a memory location read by S2.
2. *Anti-dependence:* An anti-dependence exists from S1 to S2 if S1 may read from a memory location written by S2.
3. *Output-dependence:* An output-dependence exists from S1 to S2 if S1 may write to a memory location written by S2.

Although dependence analysis is very powerful, it has its shortcomings, as shown by the simple program in Figure 1(a). In this program, there is a flow-dependence from statement S1 to S3 and from S2 to S3, and there is an anti-dependence and output-dependence from S1 to S3. There are only two statement reorderings consistent with this partial order: the original program, and the program obtained by reordering S2 and S1. In particular, the statement order shown in Figure 1(b) is not consistent with this partial order, so a compiler that relies on dependence analysis will declare that this transformation is not legal since it may not respect the semantics of the original program. It is not difficult however to verify by symbolic execution that the two programs in Figure 1 are equivalent (assuming the usual algebraic laws for numbers). If $a_{in}$ and $b_{in}$ are the values of a and b at the start of either program, the final value in a is $2*(a_{in}+b_{in})$ and the final value in b is $2*b_{in}$ in both programs. Intuitively, the constraints imposed on transformations by dependence analysis are sufficient but not strictly necessary to guarantee that transformations do not change the meaning of the program.

A more precise variation of dependence analysis is *value-based* dependence analysis [8, 17, 19]. While standard, or *memory-based* dependence analysis considers statements that touch the same memory location, value-based dependence analysis also requires that there are no intervening writes so that the statements touch the same value. Even though value-based dependence analysis is less restrictive than memory-based dependence analysis, it is still too conservative for our purposes. It is easy to verify that all dependencies in the example in Figure 1 are also value dependencies.

Although the program in Figure 1 is contrived, it illustrates an inadequacy of dependence analysis that shows up when dependence analysis is used to restructure more realistic codes like LU with pivoting. Intuitively, dependence analysis considers only the sets of locations read and written by statements; it does not assume any particular interpretation (meaning) for the operations in each statement. As our simple example shows, exploiting the semantics of these operations can lead to a richer space of program transformations.

*Symbolic analysis* is the usual way of exploiting semantics. To compare two programs for equality, we derive expressions for the outputs of these programs as functions of inputs, and attempt to prove that these expressions are equal. In principle, symbolic analysis is extremely powerful; not only does it subsume dependence analysis but it can also be used to prove equality of programs that implement very different algorithms such as sorting programs that implement quicksort and mergesort. However, for all but the simplest programs, symbolic execution and comparison is intractable. A limited kind of symbolic analysis called *value numbering* [1] and a generalization called *global value numbering* [21] are used in optimizing compilers to identify opportunities for common subexpression elimination and constant propagation, but these techniques are not useful for comparing *different* programs. Faced with this intractability, compiler-writers have settled for simple pattern-matching to identify computations in which semantic information can be exploited for restructuring. For example, all modern restructuring compilers attempt to recognize *reductions* which are statements in which a commutative and associative operation like addition is applied to the elements of an array [24]. Unfortunately, pattern matching techniques are fragile since they are easily confused by small changes to the pattern such as performing a reduction to an array location rather than to a scalar variable. Sophisticated symbolic analysis techniques for finding *generalized induction variables* have been developed by Haghighat and Polychronopoulos [11] and by Rauchwerger and Padua [20], but these techniques do not apply to the programs that we discuss.

In this paper, we propose a novel way of performing symbolic analysis of programs that we call *fractal symbolic analysis*. It is based on three ideas.

1. If the programs to be compared are too complicated for symbolic comparison, fractal symbolic analysis simplifies these programs in a way that ensures that equality of the simplified programs conversatively implies equality of the original programs.
2. In general, it is not clear how such a simplification may be accomplished, but for codes obtained by common program transformations, the appropriate simplification may be derived from the transformation.
3. This simplification process may be applied recursively until tractable programs are obtained, which is why we call this approach fractal symbolic analysis.

Our approach to simplification in this paper is inspired by Rinard and Diniz [22]. Their approach, called *commutativity analysis*, is based on the insight that a sequence of atomic operations could be executed in any permuted order (e.g., in parallel in their case) if each pair of operations can be shown to commute. While their analysis is not applicable to the program transformations we consider, we employ a more powerful variation of this idea in fractal symbolic analysis.

The rest of this paper is organized as follows. In Section 2, we introduce the highlights of our technology by discussing a small program that is a distillation of LU factorization with pivoting. We then describe our prototype implementation. In Section 3, we discuss the simplification process and derive the rules for different transformations in the literature. In Section 4, we demonstrate how we perform symbolic analysis once the programs to be compared are "simple enough". We apply this technology to automatic blocking of LU factorization with pivoting in Section 5 and show that we achieve performance comparable with that of the LAPACK library [2] on the SGI Octane. Finally, in Section 6, we discuss ongoing work.

## 2 A simple example

Figure 2(a) shows an imperfectly-nested loop nest that writes to an array A of size N. A read-only array p, whose role is similar to that of the pivot array in LU factorization, is assumed to contain integers between 1 and N such that $p(j) \geq j$. This information about array p must be provided by the programmer; in the actual LU code, it is easily deduced by the compiler, as we show in Section 5. In iteration j of the outer loop, the values in A(j) and A(p(j)) are swapped, and each element

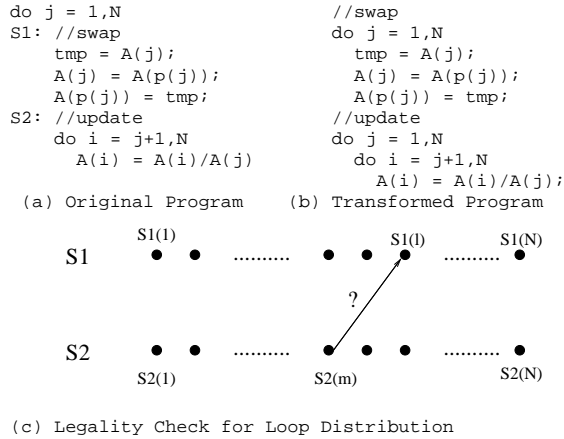

Figure 2: Loop distribution in a simple program

A(i) in the sub-array A(j+1..N) is replaced by some function of A(i) and A(j) (e.g., A(i)/A(j) in our case).

It is convenient to refer collectively to the three statements for the swap as S1, and to the update loop as S2. Each dot in Figure 2(c) represents the execution of either S1 or S2 for some iteration of the j loop. We will let S2(m) denote the execution of statement S2 in iteration m of the outer loop. Similarly, S1(l) denotes the instance of S1 for which the j loop index is l.

Figure 2(b) shows the result of *distributing* the j loop over statements S1 and S2. In the transformed program, all swaps are done before any of the updates. Since loop distribution changes the order in which operations are performed, it is not always legal. In this case, it can be seen that both programs compute the same array A provided $p(j) \geq j$.

How can a compiler reach this conclusion?

## 2.1 Dependence Analysis

As described in Section 1, dependence analysis computes a partial order between statements (in loop programs, between statement instances) by determining flow-, anti- and output-dependences. In the program of Figure 2(a), S2(1) reads and writes to location A(2) which is later read and written by S1(2). Therefore, there is a flow-dependence, an anti-dependence and an output-dependence from S2(1) to S1(2). In the transformed program, the order of execution of these two instances is reversed. Violating dependence order may change the answers produced by the program, so a compiler that uses dependence analysis will conclude conservatively that loop distribution is not legal.

## 2.2 Symbolic analysis

The most straight-forward kind of symbolic analysis performs symbolic execution of the program to compute expressions representing the values in array A at the end of execution of the original and transformed programs, and attempts to prove that these expressions are the same in both programs. This approach is relatively straightforward for basic blocks such as the programs in Figure 1, but it is not tractable for more complex programs with loops, conditionals and arrays such as the programs in Figure 2. For example, what does it mean to execute a loop when the loop bounds are symbolic expressions? We do not know any tractable way of performing symbolic evaluation and comparison even for the simple program in Figure 2, let alone LU factorization!

A more subtle symbolic analysis strategy is to use proof techniques like computational induction [10] to prove program equality. This approach is used widely in proving the correspondence of denotational and operational semantics of programs for example, but it is not clear how to use this approach for our problem. The intermediate values in array A are quite different in the two programs, and it is only at the end of execution that the values in array A in the two programs are identical, so it is difficult to think of a predicate which can be proved correct using inductive reasoning on the number of computational steps. Even if such a predicate can be designed, it is unclear how a compiler could invent it during restructuring.

## 2.3 Fractal Symbolic Analysis

These difficulties led us to a new approach to program analysis that we call *fractal symbolic analysis*. *If comparing two programs symbolically is too complicated, we simplify these programs but ensure that equality of the simplified programs implies equality of the original programs.* Intuitively, traditional symbolic analysis attempts to prove a predicate that is both necessary and sufficient to prove program equality, and gives up when the programs are too complex; in contrast, fractal symbolic analysis handles complexity by attempting to prove stronger predicates that are sufficient (but not always necessary) for program equality.

### 2.3.1 Simplifying Programs

It is not clear how to carry out this simplification in general, but in the context of this paper, we are interested only in comparing a program before and after some transformation. *This suggest that we exploit the transformation to derive the simplified programs.* To under-stand this, consider the running example of Figure 2. Imagine that loop distribution is accomplished not in a single step but *incrementally* by dragging instances of S1 in Figure 2(c) to the left over instances of S2. When all instances of S1 are scheduled before all instances of S2, we have accomplished loop distribution. At any point during this process, consider an instance S1(l) that is executed immediately *after* an instance of S2(m) (so l > m). Suppose we can show that S1(l) can be scheduled immediately *before* S2(m) without changing the result of the program. If so, we can advance the process of loop distribution one more step; repeating this argument, it is easy to see that loop distribution is legal. Therefore, if, for all $1 \leq m < l \leq N$, S2(m) and S1(l) can be executed in either order (they *commute*), loop distribution is legal. The two simplified programs we have to compare are shown in Figure 3.

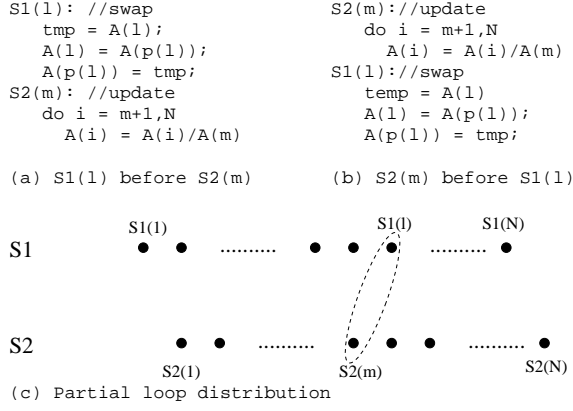

(a) S1(l) before S2(m)  (b) S2(m) before S1(l)  (c) Partial loop distribution

Figure 3: Two orders of executing statement instances

### 2.3.2 Symbolic Comparison of Simplified Programs

Our core symbolic comparison engine, described in more detail in Section 4, can analyze these simplified programs directly since they involve only statement composition and loops with no recurrences. Let $A_{in}$ denote the values in array A before the execution of the programs in Figures 3(a,b), and let $A_1$ and $A_2$ denote the values in array A after the execution of the programs in Figures 3(a) and (b) respectively. Assuming that A is the only live variable at the end of either program, we must show that $A_1 = A_2$.

Consider the program of Figure 3(a) first. Let $A_{swap}$ denote the values in array A after the swap statements have executed. Examination of the update loop shows that

$$A_1(k) = \begin{cases} 1 < m < k \leq N & \to & A_{swap}(k)/A_{swap}(m) \\ 1 \leq k \leq m \leq N & \to & A_{swap}(k) \end{cases}$$

Next, we need to express $A_{swap}$ in terms of $A_{in}$. It is easy to verify the following.

$$A_{swap}(k) = \begin{cases} k \neq l \wedge k \neq p(l) & \to & A_{in}(k) \\ k = p(l) & \to & A_{in}(l) \\ k = l & \to & A_{in}(p(l)) \end{cases}$$

By combining the expressions for $A_1$ and $A_{swap}$, we can express $A_1$ in terms of $A_{in}$ as follows.

$$A_1(k) = \begin{cases} 1 \leq k \leq m < N \\ \quad \to A_{in}(k) \\ 1 \leq m < k = l \leq N \\ \quad \to A_{in}(l)/A_{in}(m) \\ 1 \leq m < k = p(l) \leq N \\ \quad \to A_{in}(p(l))/A_{in}(m) \\ 1 \leq m < k \leq N \wedge (k \neq l) \wedge (k \neq p(l)) \\ \quad \to A_{in}(k)/A_{in}(m) \end{cases}$$

We call such an expression a *guarded symbolic expression* since it is a collection of symbolic terms with guards or predicates that specify the domain of applicability of each term.

A similar procedure can be applied to compute $A_2$ in terms of $A_{in}$ to obtain the following guarded symbolic expression for $A_2$.

$$A_2(k) = \begin{cases} 1 \leq k \leq m < N \\ \quad \to A_{in}(k) \\ 1 \leq m < k = l \leq N \\ \quad \to A_{in}(l)/A_{in}(m) \\ 1 \leq m < k = p(l) \leq N \\ \quad \to A_{in}(p(l))/A_{in}(m) \\ 1 \leq m < k \leq N \wedge (k \neq l) \wedge (k \neq p(l)) \\ \quad \to A_{in}(k)/A_{in}(m) \end{cases}$$

To compare $A_1$ with $A_2$, it is necessary to consider pairs of guards obtained by taking one guard from $A_1$ and one from $A_2$, and to verify the equality of the corresponding symbolic terms if the conjunction of the two guards is true. In this example, it is trivial to verify that $A_1$ and $A_2$ are equal. We conclude that S1(l) and S2(m) commute; therefore, the two programs in Figure 2 are themselves equal.

Note that in other programs, such as the programs in Figure 1, we may need to exploit algebraic laws to prove equality of guarded symbolic expressions. Therefore, the core symbolic comparison engine should itself be able to invoke a symbolic algebra tool like Maple [5].

### 2.3.3 Recursive Simplification

In general, the simplified programs that result by applying these rules may themselves be too complicated to

```
S1(l): //swap                S2(m,i)://update body
   tmp = A(l);                  A(i) = A(i)/A(m);
   A(l) = A(p(l));           S1(l)://swap
   A(p(l)) = tmp;               temp = A(l);
S2(m,i): //update body          A(l) = A(p(l));
   A(i) = A(i)/A(m);            A(p(l)) = tmp;
```

Figure 4: Recursive Simplification for the Running Example

be evaluated symbolically. If so, it may be necessary to apply these rules to the simplified programs recursively. This is the case for LU factorization with pivoting, as we show in Section 5. It is instructive to consider a recursive simplification of the programs in Figure 3 that eliminates the update loop. We can reorder S1(l) and S2(m) incrementally by dragging instances of the update loop S2(m,i) over S1(l). The legality of each of these incremental steps can be determined by checking if the programs in Figure 4 produce the same output for all i,m,l such that $1 \leq m < i, l \leq N$.

The two programs in Figure 4 are delightfully simple since they are just straight-line programs, but it is easy to show that they are not equal; for example, for $i = l$: $A_1(l) = A_{in}(p(l))/A(m)$ but $A_2(l) = A_{in}(p(l))$. Had we carried out the simplification of the programs in Figure 2 to this level, we would have concluded conservatively that the program transformation in Figure 2 is not legal.

This discussion highlights an important aspect of fractal symbolic analysis: successive applications of simplification produce successively stronger predicates which are less likely to be true. Therefore, the core symbolic comparison engine should be as powerful as possible so that simplification can be applied sparingly.

## 3 Fractal Symbolic Analysis

A fractal symbolic analyzer for checking legality of transformations has two components: (i) a core symbolic comparison engine for comparing programs that are "simple enough", and (ii) simplification rules for simplifying programs that are not simple enough.

To prove that a program transformation is valid, the compiler first attempts symbolic comparison (described in the next section). If the programs are not simple enough, then the compiler invokes the top-level procedure called Commute, in Figure 5 for performing fractal symbolic analysis, passing it two statements and some optional *bindings* which are constraints on free variables. These statements and bindings are obtained by the compiler from the table in Figure 6 which specifies the legality conditions for a number of common program transformations [24]. The bindings also include any constraints that the compiler can determine between free variables. The Commute procedure returns

```
Commute(stmt₁,stmt₂,bindings,live_vars) {
   if Simple(stmt₁) then
      if Simple(stmt₂)
         return Compare({stmt₁;stmt₂}, {stmt₂;stmt₁},
            bindings,live_vars)
      else
         return Commute(stmt₂,stmt₁,bindings,live_vars)
   else
      case (stmt₁) {
         ⟨stmt'₁;stmt''₁⟩ →
            return Commute(stmt'₁,stmt₂,
               bindings,live_vars) ∧ Commute(stmt''₁,stmt₂,
               bindings,live_vars)

         ⟨if pred then stmt'₁ else stmt''₁⟩ →
            return Commute(stmt'₁,stmt₂,
               bindings∥pred,live_vars) ∧ Commute(stmt''₁,
               stmt₂,bindings∥¬pred,live_vars)

         ⟨do i = l, u stmt'₁(i)⟩ →
            return Commute(stmt'₁(i),stmt₂,
               bindings∥∀i.l ≤ i ≤ u,live_vars)
      }
}
```

Figure 5: Verifying Commute Conditions

true if it can prove that the two statements commute under the constraints expressed by the bindings, and returns false otherwise. For example, to determine if the loop distribution in Figure 2 is legal, the compiler would invoke the Commute procedure with the two statements S1(l) and S2(m) in Figure 3 and the binding $1 \leq m < l \leq p(l) \leq N$.

The validity of the legality conditions in Figure 6 follows from the following result, variations of which have appeared in the literature [12].

**Lemma 1** *Let* $S = \{S_1; S_2; S_3; \ldots; S_n\}$ *be a sequence of program fragments, and let p be any permutation on S. Define* $R(p) = \{(S_i, S_j) : 1 \leq i < j \leq n \wedge p(i) > p(j)\}$ *as the set of pairs of statements reordered by p. Then, the program S is equivalent to the program* $p(S) = \{S_{p(1)}; S_{p(2)}; S_{p(3)}; \ldots; S_{p(n)}\}$ *if* $\{S_i; S_j\}$ *is equivalent to* $\{S_j; S_i\}$ *where* $(S_i, S_j) \in R(p)$.

**Proof:** The Lemma follows from induction on the cardinality of $R(p)$. If $\|R(p)\| = 0$ then clearly $S = p(S)$. Otherwise, there must exist an $i$ such that $p(i+1) < p(i)$, which implies that $(S_{p(i+1)}, S_{p(i)}) \in R(p)$. Since $S_{p(i)}$ and $S_{p(i+1)}$ commute and are adjacent in $p(S)$, transposing them gives us an equivalent program $p'(S)$ where $R(p') \subset R(p)$ and $\|R(p')\| = \|R(p)\| - 1$. By induction, $S = p(S)$. □

Intuitively, Lemma 1 allows us to reformulate the problem of checking legality of a transformation as a problem of verifying commutativity of statement instances that are reordered by that transformation. The validity of the rules given in Figure 6 follows directly from this result.

Let us now consider how the Commute procedure works. This procedure can be considered to be parameterized by a function called Compare that is the core symbolic comparison engine, invoked when both input statements are simple enough. We discuss an implementation of this function in Section 4. The boolean function Simple checks whether a statement is simple enough for Compare. If both statements are simple, the comparison engine is invoked; otherwise, the recursive simplification rules shown in Figure 7 are used to simplify the statements further. These rules are based on the syntactic structure of the two programs to be compared.

| Transformation | Legality Condition |
|---|---|
| Loop Peeling, Index Set Splitting, Skewing Inner by Outer Loop, Stripmining | true |
| Statement Reordering<br>`S1; S2;          <->   S2; S1;` | $commute(\langle S1, S2 \rangle)$ |
| Loop Fusion/Fission<br>`do i = 1,n         do i = 1,n`<br>`  S1(i);     <->     S1(i);`<br>`  S2(i);           do i = 1,n`<br>`                     S2(i);` | $commute(\langle S1(l), S2(m) \rangle : 1 <= m < l <= n)$ |
| Loop Reversal<br>`do i = 1,n         do i = n,1,-1`<br>`  S(i);      <->     S(i);` | $commute(\langle S(i), S(j) \rangle : 1 <= i < j <= n)$ |
| Loop Interchange<br>`do i = 1,n         do j = 1,m`<br>`  do j = 1,m  <->     do i = 1,n`<br>`    S(i,j);            S(i,j);` | $commute(\langle S(p,q), S(r,s) \rangle : 1 <= p < r <= n \wedge 1 <= s < q <= m)$ |
| Linear Loop Transformations<br>`do (i1,i2,...,ik)   do (i1',i2',...,ik')`<br>`  S(i1,i2,...,ik); <->  = T(i1,i2,...,ik)`<br>`                        S(i1,i2,...,ik);` | $commute(\langle S(\vec{i}), S(\vec{j}) \rangle : \vec{i} \prec \vec{j} \wedge T(\vec{i}) \succ T(\vec{j}))$ |
| Loop Tiling<br>`do i = 1,n         do I = 1,n,Bi`<br>`  do j = 1..m <->     do J = 1,m,Bj`<br>`    S(i,j);              do i = I,I+Bi-1`<br>`                           do j = J,J+Bj-1`<br>`                             S(i,j);` | $commute(\langle S(p,q), S(r,s) \rangle : (p,q) \prec (r,s) \wedge (P,Q,p,q) \succ (R,S,r,s))$ |

Figure 6: Legality Conditions for Common Program Transformations

| Transformation | Simplification Rule |
|---|---|
| Statement Sequence<br>`{S1'; S1";} B2;   <->   B2; {S1"; S1';}` | $commute(\langle S1', B2 \rangle) \wedge commute(\langle S1'', B2 \rangle)$ |
| Loop<br>`do i = l,u         B2;`<br>`  S1(i);     <->   do i = l,u`<br>`B2;                  S1(i);` | $commute(\langle S1(i), B2 \rangle : l <= i <= u)$ |
| Conditional Statement<br>`if (pred) then    B2;`<br>`  S1';             if (pred) then`<br>`else         <->    S1';`<br>`  S1";             else`<br>`B2;                 S1";` | $commute(\langle S1', B2 \rangle : pred) \wedge commute(\langle S1'', B2 \rangle : \neg pred)$ |

Figure 7: Recursive Simplification Rules

As mentioned in Section 2, the precision of fractal symbolic analysis depends on the power of the core symbolic comparison engine. Notice that the procedure in Figure 5 stops simplifying as soon as the statements being compared can be handled by the Compare procedure. A more powerful Compare procedure will result in fewer levels of simplification and potentially more accurate symbolic analysis.

## 4 Symbolic Comparison

We now describe the core symbolic comparison procedure that is invoked after simplification. As mentioned earlier, it is important for this procedure to be as powerful as is tractable so that simplification can be applied sparingly. In our work, we have found that it is sufficient if the symbolic comparison procedure can handle the following class of programs.

- Programs consist of assignment statements, for-loops and conditionals. No unstructured control flow is allowed.
- Loops do not have loop-carried dependences.
- Array indices and loop bounds are restricted to be affine functions of enclosing loop variables and symbolic constants, and predicates are restricted to be conjunctions and disjunctions of affine inequal-

$$A(\vec{k}) = \begin{cases} guard_1(\vec{k}) & \rightarrow & expression_1(\vec{k}) \\ guard_2(\vec{k}) & \rightarrow & expression_2(\vec{k}) \\ & \vdots & \\ guard_n(\vec{k}) & \rightarrow & expression_n(\vec{k}) \end{cases}$$

Figure 8: Guarded Symbolic Expressions

```
Compare(stmt₁,stmt₂,bindings,live_vars) {
   live₁ = set of live altered variables in stmt₁
   live₂ = set of live altered variables in stmt₂
   if(live₁ ≠ live₂)
      return false

   for each a(k⃗) in live₁ {
      tree₁ = Build_Expr_Tree(stmt₁,a(k⃗),∅)
      tree₂ = Build_Expr_Tree(stmt₂,a(k⃗),∅)

      gse₁ = Normalize_GSE(tree₁,bindings)
      gse₂ = Normalize_GSE(tree₂,bindings)

      if(¬ Compare_GSEs(gse₁,gse₂)
         return false
   }
   return true
}
```

Figure 9: Comparision of Simple Programs

ities.

The important constraint is the second one. Although a loop may write to a section of an array that is potentially unbounded at compile-time, at most one iteration may effect the value of any given location in an array. This ensures that the symbolic value of a given element of the array can be expressed finitely. Techniques such as scalar expansion [24] should be used to aggressively eliminate loop-carried dependences.

We can then summarize the unbounded set of expressions for the values in an entire array with a finite expression called a *guarded symbolic expression* (or *GSE* for short) which contains symbolic expressions that hold for affinely constrained portions of the array as shown in Figure 8. Section 2 contains a number of examples of guarded symbolic expressions.

Figure 9 provides a high-level overview of our symbolic comparison algorithm. We consider each altered scalar or array variable in the two programs being compared. Note that we only need to consider *live* variables [1]. If the GSE's corresponding to each live altered variable are equal, the two programs are declared to be equal. We now describe how GSE's are constructed and compared.

### 4.1 Generation of Conditional Expression Trees

A guarded symbolic expression is essentially a description of the effect of a program on an array. As an in-

```
Build_Expr_Tree(stmt,tree,bindings) {
   case (tree) {
      Op(op,tree₁,...,treeₙ) →
         return Op(op,
            Build_Expr_Tree(stmt,tree₁,bindings),...,
            Build_Expr_Tree(stmt,tree₁,bindings))

      Cond(pred,treeₜ,tree_f) →
         return Cond(pred,
            Build_Expr_Tree(stmt,treeₜ,bindings),
            Build_Expr_Tree(stmt,tree_f,bindings))

      A(k⃗) →
         case (stmt) {
            ⟨A'(T · i⃗ + c) = tree₁(i⃗)⟩ →
               if (A = A') then
                  return Cond(bindings∥k⃗ = T · i⃗ + c,
                     tree₁(T⁻¹ · (k⃗ − c)),A(k⃗))
               else
                  return A(k⃗)

            ⟨stmt₁;stmt₂⟩ →
               return Build_Expr_Tree(stmt₁,
                  Build_Expr_Tree(stmt₂,tree,bindings),
                  bindings)

            ⟨if pred then stmt₁ else stmt₂⟩ →
               return Cond(bindings∥pred,
                  Build_Expr_Tree(stmt₁,tree,bindings),
                  Build_Expr_Tree(stmt₂,tree,bindings))

            ⟨do iₖ = lₖ, uₖ stmt₁⟩ →
               return Build_Expr_Tree(stmt₁,tree,
                  bindings∥∃iₖ.lₖ ≤ iₖ ≤ uₖ)
         }
   }
}
```

Figure 10: Expression Tree Generation

```
Cond(∃j.m + 1 ≤ j ≤ N ∧ k = j,
   Op(/,
      Cond(k = p(l),A(l),Cond(k = l,A(p(l)),A(k))),
      Cond(m = p(l),A(l),Cond(m = l,A(p(l)),A(m)))),
   Cond(k = p(l),A(l),Cond(k = l,A(p(l)),A(k))))
```

(i) Conditional Expression Tree for Figure 2(a)

```
Cond(k = p(l),
   Cond(∃j.m + 1 ≤ j ≤ N ∧ l = j,
      Op(/,A(l),A(m)),
      A(l)),
   Cond(k = l,
      Cond(∃j.m + 1 ≤ j ≤ N ∧ p(l) = j,
         Op(/,A(p(l)),A(m)),
         A(p(l))),
      Cond(∃j.m + 1 ≤ j ≤ N ∧ k = j,
         Op(/,A(k),A(m)),
         A(k))))
```

(ii) Conditional Expression Tree for Figure 2(b)

Figure 11: Two Examples of Conditional Expression Trees

termediate step towards the construction of this description, we build a symbolic representation of the program that we call a *conditional expression tree*. A conditional expression tree may be viewed as a functional representation of the portion of the program required to compute the final values of a given array. Figure 10 shows an algorithm to generate such trees. This algorithm processes

```
Factor(tree) {
  case (tree) {
    Op(op,tree_1,...,Cond(pred,tree_t_i,tree_f_i),...,tree_n) →
      return Factor(Cond(pred,
              Op(op,tree_1,...,tree_t_i,...,tree_n),
              Op(op,tree_1,...,tree_f_i,...,tree_n)))

    Op(op,tree_1,...,tree_n) →
      return Op(op,Factor(tree_1),...,,Factor(tree_n))

    Cond(pred,tree_t,tree_f) →
      return Cond(pred,Factor(tree_t),Factor(tree_f))

    A(k⃗) →
      return A(k⃗)
  }
}

Build_GSE(tree,guard)
  if (guard) then
    case (tree) {
      Cond(pred,tree_t,tree_f) →
        return Build_GSE(tree_t,guard ∧ pred) ⋃
               Build_GSE(tree_f,guard ∧ ¬pred)

      Op(op,tree_1,...,tree_n) →
      A(k⃗) →
        return {(guard, expr)}
    }
  else
    return ∅
}

Normalize_GSE(tree,bindings) {
  return Build_GSE(Factor(tree),bindings)
}
```

Figure 12: From Expression Trees to GSE's

the statements of a program in reverse order, determining at each step the tree corresponding to relevant output data in terms of input data and linking these together to produce the final result. Figure 11 illustrates the conditional expression trees generated from the programs in Figure 2.

### 4.2 Normalization to Guarded Symbolic Expressions

The conditional expression trees generated above contain a mix of conditions predicated by affine constraints on one hand and symbolic expressions on the other. To convert these to guarded symbolic expressions, we need to separate the two. We accomplish this by factoring the affine constraints outside of the symbolic operations by repeated application of the following transformation.

```
Op(op,tree_1,...,Cond(pred,tree_t_i,tree_f_i),...,tree_n)
                        ⇓
    Cond(pred,Op(op,tree_1,...,tree_t_i,...,tree_n),
              Op(op,tree_1,...,tree_f_i,...,tree_n))
```

At this point, the guards are generated by combining the predicates at the top of the factored expression tree, and the corresponding symbolic expressions are simply taken from the subtrees beneath these predicates. This is shown in Figure 12.

```
Compare_GSEs(gse_1,gse_2) {
  for each (guard_1,expr_1) in gse_1 {
    for each (guard_2,expr_2) in gse_2 {
      if (guard_1 ∧ guard_2 ≠ false)
        if (expr_1 ≠ expr_2)
          return false
    }
  }
  return true
}
```

Figure 13: Comparision of GSE's

### 4.3 Comparison of Guarded Symbolic Expressions

Finally, Figure 13 illustrates the comparison of two guarded symbolic expressions. There are two steps to this comparison. First, we must compare each pair of affine guards of the two guarded symbolic expressions. Second, for any two guards that potentially intersect, we must compare the corresponding symbolic expressions. If all such symbolic expression match, then the guarded symbolic expressions are declared to be equal. The validity of this conclusion follows from the following argument. Each guard specifies some region of the index space of the array in question, and the union of these regions in a guarded symbolic expression is equal to the entire index space of that array. If the values in the two guarded symbolic expressions are identical whenever their guards intersect, the two array values are obviously equal.

For comparison of affine guards, we may employ an integer programming tool such as the Omega Library [18], which we have chosen for our implementation. If our tool can automatically verify that a pair of affine guards do not intersect, there is no need for further comparison.

For comparison of symbolic expressions, we currently test for syntactic equality. This is sufficient both for our simple example and, as we shall see in Section 5, for LU factorization. It would be easy to use Maple or some other symbolic algebra tool for this test if we wished to exploit algebraic properties of numbers.

## 5 Blocking LU with pivoting

Fractal symbolic analysis was developed for use in an ongoing project on optimizing the cache behavior of dense numerical linear algebra programs. LU factorization with partial pivoting is a key routine in this application area since it is used to solve systems of linear equations of the form Ax = b. Figure 14 shows the canonical *right-looking* version of LU factorization with pivoting that appears in the literature [9]. In iteration j of the outer loop, computations are performed on column j of the matrix A, and a portion of the matrix to the *right* of this column is updated. The `i` and `k` loops in the update step can be interchanged, giving two versions of right-looking LU factorization. A rather different version of LU factorization is called *left-looking* LU factorization; intuitively, this version delays the updates made by the right-looking version to a column till it is time to compute with that column.

Cache-optimized versions of LU factorization can be found in the LAPACK library [2]. These *blocked codes* are too complex to be reproduced here, but they perform much better than the *point version* shown in Figure 14. Figure 19 shows the performance of the point version and two blocked versions on an SGI Octane[1]. One blocked version was obtained from the Netlib repository[2], and it is a portable blocked LU that calls BLAS [9] routines tuned for the Octane to perform key operations like matrix multiplication. The second blocked version was written at SGI for the Octane. Figure 19 shows that the performance of the point version degrades to about 70 MFlops for large matrix sizes; in contrast, the Netlib blocked code obtains about 425 MFlops, while the SGI blocked code obtains about 475 MFlops.

LU factorization with pivoting poses a number of challenges for compiler writers.

1. Given point-wise LU factorization with pivoting, can a compiler automatically generate a cache-optimized version by blocking the code? If so, how does the performance of the compiler-optimized code compare with that of hand-blocked code?
2. Modern restructuring compilers can transform one version of right-looking LU factorization to the other automatically by interchanging the two loops of the update step. Can a compiler transform right-looking LU to left-looking LU and vice versa?

Fractal symbolic analysis is crucial to address both these challenges. For lack of space, we discuss only the problem of blocking.

### 5.1 Automatic Blocking of LU factorization

To obtain code competitive with LAPACK code, Carr and Lehoucq suggest carrying out the following sequence of transformations [4].

1. Stripmine the outer loop to operate on block-columns.
2. Index-set-split the expensive update operation to

[1] This 300MHz machine has a 2MB L2 cache, and an R12K processor. All compiled code was generated using the SGI MIPSpro f77 compiler with flags: -O3 -n32 -mips4.
[2] http://www.netlib.org

```
do j = 1, N
  // Pick the pivot
  p(j) = j
  do i = j+1, N
    if abs(A(i,j)) > abs(A(p(j),j))
      p(j) = i

  // Swap rows
  do k = 1, N
    tmp = A(j,k)
    A(j,k) = A(p(j),k)
    A(p(j),k) = tmp

  // Scale current column
  do i = j+1, N
    A(i,j) = A(i,j) / A(j,j)

  // Update portion of matrix
  // to right of column j
  do k = j+1, N
    do i = j+1, N
      A(i,k) = A(i,k) - A(i,j)*A(j,k)
```

Figure 14: LU Factorization with Pivoting

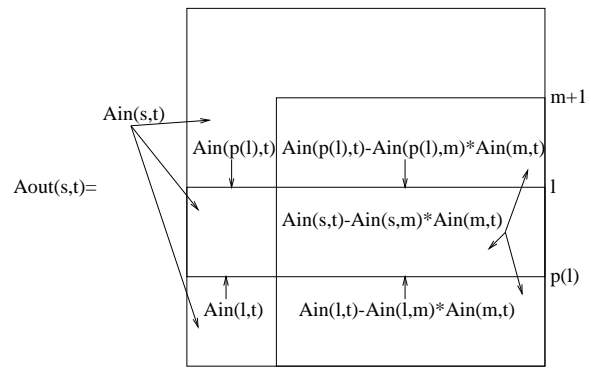

Figure 15: Regions and Expressions for Simplified LU

separate computation outside the current block-column from that inside.
3. Distribute the inner of the stripmined loops to isolate the out-of-column update.
4. Tile the out-of-column update.

The first of two steps, stripmining and index-set-splitting, are trivially legal as they do not reorder any computation. The next step, loop distribution, is not necessarily legal. If this legality is checked using dependence analysis, the compiler declares the distribution illegal if there is a dependence from an iteration `B2(m)` to an iteration `B1(l)` where `l > m`. In fact, such a dependence exists in our program; for example, both `B2(j)` and `B1(j+1)` read and write to `A(m+1,jB+B..N)`. *Therefore, a compiler that relies on dependence analysis cannot block LU with pivoting using the transformation strategy of Carr and Lehoucq.*

Carr and Lehoucq suggest that a compiler may be endowed with application-specific information to recognize the swap and update operations in LU factorization, and to realize that they can be legally interchanged.

```
   do jB = 1, N, B                                     do jB = 1, N, B
     do j = jB, jB+B-1                                   do j = jB, jB+B-1
B1(j):                                          B1(j):
        // Pick the pivot                                   // Pick the pivot
        p(j) = j                                            p(j) = j
        do i = j+1, N                                       do i = j+1, N
          if abs(A(i,j)) > abs(A(p(j),j))                     if abs(A(i,j)) > abs(A(p(j),j))
            p(j) = i                                            p(j) = i

        // Swap rows                                        // Swap rows
        do k = 1, N                                         do k = 1, N
          tmp = A(j,k)                                        tmp = A(j,k)
          A(j,k) = A(p(j),k)                                  A(j,k) = A(p(j),k)
          A(p(j),k) = tmp                                     A(p(j),k) = tmp

        // Scale column                                     // Scale column
        do i = j+1, N                                       do i = j+1, N
          A(i,j) = A(i,j) / A(j,j)                            A(i,j) = A(i,j) / A(j,j)

        // In-Column Update                                 // In-Column Update
        do k = j+1, jB+B-1                                  do k = j+1, jB+B-1
          do i = j+1, N                                       do i = j+1, N
            A(i,k) = A(i,k) - A(i,j)*A(j,k)                     A(i,k) = A(i,k) - A(i,j)*A(j,k)

B2(j):                                              // Distributed Loop
        // Right-Looking Update                             do j = jB, jB+B-1
        do k = jB+B, N                              B2(j):
          do i = j+1, N                                     // Right-Looking Update
            A(i,k) = A(i,k) - A(i,j)*A(j,k)                 do k = jB+B-1, N
                                                              do i = j+1, N
                                                                A(i,k) = A(i,k) - A(i,j)*A(j,k)
```

        (a) Before Loop Distribution                                (b) After Loop Distribution

Figure 16: LU Factorization: Distribution Step

Fractal symbolic analysis is a general-purpose technique that makes this unnecessary.

The rules for legality of loop distribution in Fig 6 require that B1(l) commute with B2(m) where $jB \leq m < l \leq jB + B - 1$, as shown in Figure 17. The compiler invokes the Commute method in Figure 5 with these parameters. However, these simpler programs are not "simple enough"; the loop that computes the pivot in B1.b(l) is a recurrence that cannot be handled by our core symbolic comparison engine, as we discussed in Section 4. Therefore, these programs are simplified again using the rule for statement sequences in Figure 7. This requires the compiler to test whether B2(m) commutes with the five subblocks in B1(l). With the exception of B1.c(l), the data touched by each of the subblocks of B1(l) is disjoint from the data touched by B2(m). Therefore, the compiler deduces that these subblocks commute with B2(m) (a small detail is that the analysis of whether B1.b(l) commutes with B2(m) requires an additional step of simplification to eliminate the recurrence in B1.b(l)).

The difficult part of this process is to demonstrate that B1.c(l) and B2(m) commute as shown in Figure 18. At this point, these programs are "simple enough", and the Compare method in Figure 9 is invoked to establish equality of the simplified programs. In fact, they are quite similar to those in our simple example and guarded expressions are generated in the same fashion as discussed in Section 4. The only live, altered variable in either program is the array A, and the Compare method generates guarded symbolic expressions for A from each program. Both GSE's generated from Figure 18 contain six guarded regions, correlating directly to the picture in Figure 15. To prove that the GSE's are actually equivalent, Compare_GSEs is invoked to test the 36 pairwise intersections, and the Omega library [18] is used to test the satisfiability. Only six intersections are non-empty, and the corresponding symbolic expressions are syntactically identical in each case. Thus, the compiler is able to demonstrate the equality of the simplified programs and, therefore, the programs in Figure 16.

Note that the programs are only equivalent given that $p(j) \geq j$. Techniques such as value propagation [16, 7] have been developed to perform this type of analysis for indirect array accesses to more accurately compute dependences. It is clear that this information may easily be inferred from the pivot computation in B1.a and B1.b. This information should be passed by the compiler as bindings to the method Commute along with the legality conditions in Figure 6.

With this information, our implementation of fractal symbolic analysis is able to automatically establish the legality of the loop distribution transformation in Figure 16. Although the algorithm is exponential (e.g., the Omega library itself is exponential), in practice it is reasonably fast. For this example, our implementa-

```
B2(m):   do k = jB+B, N
           do i = m+1, N
             A(i,k) = A(i,m) - A(i,m)*A(m,k)

B1.a(l): p(l) = l

B1.b(l): do i = l+1, N
           if abs(A(i,l)) > abs(A(p(l),l))
             p(l) = i

B1.c(l): do k = 1, N
           tmp = A(l,k)
           A(l,k) = A(p(l),k)
           A(p(l),k) = tmp

B1.d(l): do i = l+1, N
           A(i,l) = A(i,l) / A(l,l)

B1.e(l): do k = l+1, jB+B-1
           do i = l+1, N
             A(i,k) = A(i,k) - A(i,l)*A(l,k)
```

(a) `B2(m); B1(l)`

```
B1.a(l): p(l) = l

B1.b(l): do i = l+1, N
           if abs(A(i,l)) > abs(A(p(l),l))
             p(l) = i

B1.c(l): do k = 1, N
           tmp = A(l,k)
           A(l,k) = A(p(l),k)
           A(p(l),k) = tmp

B1.d(l): do i = l+1, N
           A(i,l) = A(i,l) / A(l,l)

B1.e(l): do k = l+1, jB+B-1
           do i = l+1, N
             A(i,k) = A(i,k) - A(i,l)*A(l,k)

B2(m):   do k = jB+B, N
           do i = m+1, N
             A(i,k) = A(i,m) - A(i,m)*A(m,k)
```

(b) `B1(l); B2(m)`

Figure 17: Simplified Comparison #1

```
B2(m):   do k = jB+B, N
           do i = m+1, N
             A(i,k) = A(i,m) - A(i,m)*A(m,k)

B1.c(l): do k = 1, N
           tmp = A(l,k)
           A(l,k) = A(p(l),k)
           A(p(l),k) = tmp
```

(a) `B2(m); B1.c(l)`

```
B1.c(l): do k = 1, N
           tmp = A(l,k)
           A(l,k) = A(p(l),k)
           A(p(l),k) = tmp

B2(m):   do k = jB+B, N
           do i = m+1, N
             A(i,k) = A(i,m) - A(i,m)*A(m,k)
```

(b) `B1.c(l); B2(m)`

Figure 18: Simplified Comparison #2

tion, prototyped in Caml-Light [14], took slightly less than one second, much faster than the corresponding reduction in execution time of a single application of LU for medium and large size matrices. Most of the analysis time is spent on the construction and comparison of guarded symbolic expressions. We are pursuing several strategies to improve the analysis time.

## 5.2 Experimental Results

To study the effects of automatic blocking, we ran the SGI compiler on the LU code in Figure 16, after loop distribution is performed. Given just this slightly transformed code, the SGI compiler is now able to produce significantly faster code that does not degrade as matrices exceed the cache. Once the loop is distributed, the compiler is able to automatically tile the right-looking update (B2) and essentially accomplish the last Carr/Lehoucq step listed above.

Nevertheless, this code, at 200 MFlops, is still a factor of two slower than the LAPACK codes. Further experimentation found the remaining performance gap due the compiler's suboptimal treatment of the right-looking update computation. Although, the SGI compiler is able to now block the update, we surmise that it may have been confused by the partially triangular loop bounds of the update. When we index-set split the i loop by hand to separate the triangular and rectangular portions of the update, the compiler generated substantially faster code achieving over 300 MFlops. Finally, we note that if we replace the triangular and rectangular portions of the update with the corresponding BLAS-3 calls (DTRSM and DGEMM) used in LAPACK, the resulting code achieves nearly 400 MFlops and is within 10% of Netlib LAPACK and 20% of the best hand-optimized code. Thus, with the ability to isolate the update, we believe that compilers should be able to nearly match the LAPACK as their ability to match the performance of BLAS on perfectly nested codes improves.

## 6 Conclusions and Future Work

In this paper, we presented a new analysis technique called fractal symbolic analysis for proving the validity of program transformations.

We are currently exploring different options in the design of the fractal symbolic analyzer. In principle,

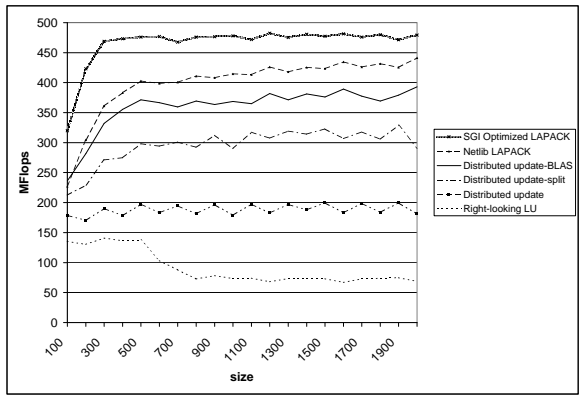

Figure 19: Transforming LU with Pivoting

the symbolic comparison engine can be extended to recognize and summarize reductions involving associative arithmetic operations like addition and multiplication, perhaps using the techniques of Haghighat and Polychronopoulos [11].

At present, we only perform syntactic comparisons of the symbolic expressions in guarded symbolic expressions. A symbolic algebra tool like Maple [5] will add to the power of the comparison engine, and this power will be useful once we recognize and summarize reductions. These enhancements might eliminate the need for recursive simplification in some programs, but we do not yet have any applications where this additional power is needed.

The algorithm for generating guarded symbolic expressions in Section 2.3.2 is reminiscent of backward slicing [23] which is a technique that isolates the portion of a program that may affect the value of a variable at some point in the program. Our algorithm is simpler than the usual algorithms for backward slicing since the programs it must deal with have been simplified beforehand by recursive simplification, an operation that has no analog in backward slicing. A similar statement can be made about the computation of *last-write* information [6].

Finally, we note that dependence information for loops can be represented abstractly using dependence vectors, cones, polyhedra etc. These representations have been exploited to *synthesize* transformation sequences [3, 13, 15]. At present, we do not know suitable representations for the results of fractal symbolic analysis, nor do we know how to synthesize transformation sequences from such information.